\documentclass[12pt]{article}
\usepackage{amssymb,epsfig}

\renewcommand{\thefootnote}{\fnsymbol{footnote}}

\begin{document}

\title{
\begin{flushright}
\ \\*[-80pt] 
\begin{minipage}{0.25\linewidth}
\normalsize
YITP-07-37 \\
KAIST-TH 2007/04\\
KUNS-2081 \\*[50pt]
\end{minipage}
\end{flushright}
{\Large \bf 
TeV scale partial mirage unification 
\\ and neutralino dark matter
\\*[20pt]}}

\author{Hiroyuki~Abe$^{1,}$\footnote{
E-mail address: abe@yukawa.kyoto-u.ac.jp}, \ 
Yeong~Gyun~Kim$^{2,3}$\footnote{
E-mail address: ygkim@muon.kaist.ac.kr}, \ 
Tatsuo~Kobayashi$^{4,}$\footnote{
E-mail address: kobayash@gauge.scphys.kyoto-u.ac.jp}, \\ 
 and \ Yasuhiro~Shimizu$^{2,}$\footnote{
E-mail address: shimizu@muon.kaist.ac.kr}\\*[20pt]
$^1${\it \normalsize 
Yukawa Institute for Theoretical Physics, Kyoto University, 
Kyoto 606-8502, Japan} \\
$^2${\it \normalsize 
Department of Physics, Korea Advanced Institute of Science
and Technology,} \\ {\it \normalsize Daejeon 305-701, Korea} \\
$^3${\it \normalsize 
ARCSEC, Sejong University, 
Seoul 143-747, Korea} \\
$^4${\it \normalsize 
Department of Physics, Kyoto University, 
Kyoto 606-8502, Japan} \\*[50pt]}

\date{
\centerline{\small \bf Abstract}
\begin{minipage}{0.9\linewidth}
\medskip 
\medskip 
\small
We study the TeV scale partial mirage unification scenario, 
where the gluino and wino masses are degenerate around 
a TeV scale, but the bino mass is not degenerate.
This scenario has phenomenologically interesting aspects.
First, because of the degeneracy between the 
gluino and wino masses, this scenario does not have 
the little hierarchy problem, that is, the higgisino mass is 
around 150 GeV.
The lightest superparticle is a mixture of the bino and higgsino, 
and can lead to a right amount of thermal relic density as a dark matter 
candidate.  
\end{minipage}
}

\begin{titlepage}
\maketitle
\thispagestyle{empty}
\end{titlepage}


\renewcommand{\thefootnote}{\arabic{footnote}}
\setcounter{footnote}{0}

\section{Introduction}
\label{sec:intro}
Supersymmetric extension of the standard model (SM) is one of the 
most promising candidates for a new physics at the TeV scale. 
In particular, the minimal supersymmetric standard model (MSSM) 
is interesting from the viewpoint of its minimality.
The MSSM has several attractive aspects.
The MSSM realizes the unification of three gauge couplings
at the grand unified 
theory (GUT) scale $M_{GUT} \sim 2 \times 10^{16}$ GeV.
Supersymmetry can stabilize the huge hierarchy 
between the weak scale and the GUT/Planck scale.
Supersymmetric models with R-parity have a good candidate for dark matter, 
that is, the lightest superparticle (LSP).

However, these attractive aspects are not perfectly 
satisfying.
First of all, there is still a fine-tuning problem as follows.
By minimizing the Higgs scalar potential, 
the $Z$ boson mass is obtained as 
\begin{eqnarray}
\frac{1}{2}M_Z^2 &\sim& -\mu^2(M_Z)-m_{H_u}^2(M_Z), 
\nonumber
\end{eqnarray}
where  $\mu$ is the SUSY mass of up- and down-sector Higgs fields
and $m_{H_u}$ is the soft SUSY breaking mass for 
the up-sector Higgs field.
Thus, natural values of $|\mu^2|$ and $|m_{H_u}^2|$ would be 
of $O(M_Z^2)$.
Otherwise, we need fine-tuning between $|\mu^2|$ and $|m_{H_u}^2|$ 
to cancel them and to lead to $M_Z$.
The soft mass $m_{H_u}$  receives a large radiative correction 
between the weak scale and the cut-off scale $\Lambda$,
\begin{eqnarray}
\Delta m_{H_u}^2 &\sim& 
-\frac{3y_t^2}{4\pi^2} m_{\tilde{t}}^2 
\ln \frac{\Lambda}{m_{\tilde{t}}}, 
\nonumber
\end{eqnarray}
where $y_t$ is the top Yukawa coupling and 
$m_{\tilde{t}}$ is the stop mass.
The cut-off scale may be the GUT scale or Planck scale, 
and we may have $\Delta m_{H_u}^2 \sim -2 m_{\tilde{t}}^2 $ 
or $-3 m_{\tilde{t}}^2$.
On the other hand, the theoretical upper bound for 
the lightest CP-even Higgs mass is obtained as \cite{Okada:1990gg}
\begin{eqnarray}
m_h^2 &\le& 
M_Z^2  + \frac{3 m_t^4}{4 \pi^2 v^2} 
\ln \frac{m_{\tilde{t}}^2}{m_t^2} + \cdots.
\nonumber
\end{eqnarray} 
The experimental bound $m_h \ge 114.4$ GeV requires 
$m_{\tilde{t}} \gtrsim 500$ GeV.
This value of $m_{\tilde{t}}$ leads to a quite large 
correction $\Delta m_{H_u}^2$.
Hence, in order to realize $M_Z=91$ GeV, 
we need a few percent fine-tuning 
of the SUSY mass $\mu$ and the soft SUSY breaking mass 
$m_{H_u}$ at the GUT scale, although these two masses are, 
in general, independent parameters.
This fine-tuning problem is sometimes called as 
the little-hierarchy problem between the weak scale and 
a TeV scale ~\cite{Barbieri:1987fn}.

Several types of models have been proposed 
to solve the little hierarchy problem.
Among them, the TeV scale mirage mediation 
\cite{Choi:2005hd,Kitano:2005wc,Choi:2006xb} is 
one of most interesting scenarios, 
because the field content in the visible sector 
is the same as one of the MSSM.
In the mirage mediation, the modulus 
mediation and anomaly mediation \cite{anomaly} are comparable 
\cite{Choi:2005ge,Choi:2005uz}, and such situation 
can be realized in the KKLT-type 
of moduli stabilization \cite{Kachru:2003aw}.
One of interesting aspects in the mirage mediation 
is that the anomaly mediation effect and 
renormalization group (RG) effects cancel each other 
at the so-called mirage scale $M_{\rm mir}$.
That is, soft SUSY breaking terms at $M_{\rm mir}$ 
appear equivalent to the pure modulus mediation, 
although there is no physical threshold at $M_{\rm mir}$.
Therefore, the model with $M_{\rm mir}=O({\rm TeV})$, 
i.e., the TeV scale mirage model, is interesting 
as a solution of the little hierarchy problem.
In the TeV scale mirage model, the superparticle 
spectrum derived from the pure modulus mediation 
appears at the TeV scale through the cancellation 
between the anomaly mediation and RG effects.
In particular, the modulus mediation leading to 
\begin{equation}
|m_{H_u}| \sim \mu \sim M_Z, \qquad m_{\tilde t}=O(1){\rm TeV}
\end{equation}
is interesting.
Indeed, concrete models realizing the above spectrum 
have been studied in Ref.~\cite{Choi:2005hd,Choi:2006xb}.
In those models, gauge kinetic functions for 
three MSSM vector multiplets are universal 
and three gaugino masses are universal at $M_{\rm mir}$.
In addition, the universal gaugino mass is of $O(1)$ TeV.

One of interesting aspect in SUSY models is that 
they have a good candidate for the dark matter as 
the LSP.
In the above TeV scale mirage scenario, 
the value of $\mu$ is of $O(M_Z)$ 
to avoid fine-tuning, 
while the gaugino masses are universal around a TeV scale and 
it is of $O(1)$ TeV.
Thus, the LSP is higgsino-like in the TeV scale mirage scenario.
In this case, the thermal relic density of the LSP is 
much lower than cosmological observation \cite{Choi:2006im}.

Recently, the bottom-up analysis \cite{Abe:2007kf}
showed that the degeneracy between the wino and gluino masses 
is most important to avoid fine-tuning in the Higgs sector, 
but the bino mass can vary with keeping the same 
degree of fine-tuning.
When the bino mass varies, several phenomenological aspects 
would change.
The LSP is a mixture of the higgsino and bino, 
and its thermal relic density would be totally different from 
one of higgsino-like LSP.
Hence, in this paper we study the TeV scale partial 
mirage unification, where the wino and gluino masses 
are degenerate around $O(1)$ TeV, but the bino mass 
is different.
We study phenomenological aspects of this scenario, 
in particular the thermal relic density and direct detection possibility
for the neutralino LSP.\footnote{See
  Ref.\cite{naturaldm,Dermisek:2006ey} 
for other studies on
a connection between naturalness of electroweak symmetry breaking and dark
matter phenomenology.}

Several authors have investigated phenomenological and 
cosmological aspects of mirage mediation 
\cite{endo05,falkowski05,yama,baer,kitano-lhc,Choi:2006im,kawagoe,cho}.
In particular, it has been noticed that moduli decay in the early
universe can produce so many gravitinos and neutralino LSPs that 
successful Big Bang nucleosynthesis might be ruined and/or too large dark
matter abundance would be obtained \cite{yama}. 
A possible way out of the cosmological moduli problem is to dilute the primordial moduli and
the subsequently produced gravitinos and LSPs, through some mechanism
such as the thermal inflation \cite{stewart}. In this work, we assume that 
such a mechanism is realized and the neutralino dark matter is generated
through the conventional thermal production mechanism.

This paper is organized as follows.
In section 2, we study a concrete model leading to 
the TeV scale partial mirage unification.
In section 3, we study phenomenological aspects of 
our scenario, in particular, the thermal relic 
density of the LSP.
Section 4 is devoted to conclusion and discussion.

\section{TeV scale partial mirage unification}

\subsection{Moduli stabilization 
in the generalized KKLT scenario}

Indeed, our model is quite similar to the model 
for the TeV scale mirage \cite{Choi:2006xb}, 
which is a generalization of the KKLT scenario for 
moduli stabilization \cite{Kachru:2003aw,Abe:2005rx,Choi:2006bh}.

We consider the IIB string model with the dilaton $S$,  
a single K\"ahler modulus $T$ and complex 
structure moduli $Z_\alpha$.
First, we assume that the dilaton $S$ and complex structure moduli 
$Z_\alpha$ are stabilized by the flux-induced superpotential 
$W_{\rm flux}(S,Z_\alpha)$ \cite{Giddings:2001yu}, that is, they have 
heavy masses of $O(M_P)$, where $M_P$ is the Planck scale.
At this stage, the K\"ahler modulus $T$ is not stabilized.
To stabilize $T$, we introduce a $T$-dependent non-perturbative effect
in the superpotential.
In the original KKLT model, a simple term is considered as 
\begin{equation}
W_{\rm np} = A e^{-aT},
\end{equation}
where $A = O(M_P^3)$ and $a$ is a constant.
Here and hereafter we use the unit, where $M_P=1$.
Such non-perturbative effect can be generated by a gaugino 
condensation of the hidden gauge sector on D7 branes, 
whose gauge kinetic function is proportional to $T$.
In general, the gauge kinetic function is a linear 
combination of $S$ and $T$ as, 
\begin{equation}
f_a=k_a T + \ell_a S,
\end{equation}
e.g. on magnetized D-branes, where 
$k_a$ and $\ell_a$ are rational numbers \cite{Lust:2004cx}.
The gaugino condensation in the hidden gauge sector may generate 
a non-perturbative term 
like $W_{\rm np} \sim e^{-8 \pi^2 (k_h T + \ell_h S)}$.
Thus, we consider the superpotential \cite{Abe:2005rx}
\begin{equation}
W = \langle W_{\rm flux}\rangle - A_h e^{-8 \pi^2 (k_h T + \ell_h S)},
\end{equation}
where $A_h=O(1)$ and $8 \pi^2 k_h =O(10)$.
In the second term of the right hand side, the dilaton $S$ is 
replaced by its vacuum expectation value (VEV) $S_0$, 
because $S$ is assumed to be stabilized with a mass of $O(M_P)$ 
by the flux-induced superpotential $W_{\rm flux}$.\footnote{
We can replace $S$ by its VEV only when the  flux-induced
superpotential $W_{\rm flux}$ includes its supersymmetric mass, 
which is heavier than the mass of $T$ and the gravitino mass.
Otherwise, such analysis is not 
valid \cite{Choi:2004sx,deAlwis:2005tf}.}
With this superpotential and the K\"ahler potential, 
\begin{equation}
K_0=-3\ln (T + T^*),
\end{equation}
we can write the scalar potential, 
\begin{equation}
V_F = e^{K_0}\left[ K_0^{T T^*}|D_TW|^2 -3|W|^2 \right],
\end{equation}
where
\begin{equation}
D_TW \equiv (\partial_T K_0 )W + \partial_T W.
\end{equation}
The K\"ahler modulus $T$ is stabilized at the SUSY point $D_TW =0$, 
where we can estimate $\langle W \rangle \approx \langle W_{\rm
  flux}\rangle$ because $8 \pi^2 k_h =O(10)$.
At this SUSY point, 
the vacuum energy is negative,
\begin{equation}
V_F = -3m^2_{3/2},
\end{equation}
where $m_{3/2}$ denotes the gravitino mass, $m_{3/2}=e^{K_0/2}W$.
The modulus $T$ has a mass of $O(8 \pi^2 m_{3/2})$, which is 
much larger than the gravitino mass.

To obtain a de Sitter (Minkowski) vacuum, we add the uplifting 
potential, 
\begin{equation}
V_{\rm lift} = e^{2K_0/3}{\cal P}_{\rm lift}.
\end{equation}
Such potential can be generated by putting anti D3-brane 
at a tip of warp throat  \cite{Kachru:2003aw}, 
and the warp factor leads to a suppressed value of 
${\cal P}_{\rm  lift}$ \cite{Giddings:2001yu}, 
\begin{equation}
{\cal P}_{\rm lift} \sim e^{-32\pi^2K{\rm Re}(S_0)/3M},
\end{equation}
where $K$ and $M$ are integer-valued NS and R 3-form fluxes.
We tune our parameters to realize almost vanishing 
vacuum energy, i.e. $V_F+ V_{\rm lift} \approx 0$.
Since $V_F \approx -3 e^{K_0}|W_{\rm flux}|^2$,  the above fine-tuning 
requires $|W_{\rm flux}|^2 \sim {\cal P}_{\rm lift} \sim
e^{-32\pi^2K{\rm Re}(S_0)/3M}$.
Hence, we can parameterize $W_{\rm flux}$ as 
\begin{equation}
W_{\rm flux} = A_0 e^{-8 \pi^2 \ell_0 S_0},
\end{equation}
where $\ell_0$ is a rational number and $A_0 =O(1)$.
We consider the low-energy SUSY breaking, i.e. 
$m_{3/2} = O(10)$ TeV.
That requires 
\begin{equation}
8 \pi^2 \ell_0 {\rm Re}(S_0) \simeq \ln (M_P/m_{3/2}) \sim 
4 \pi^2.
\end{equation}
At the minimum of $V_F + V_{\rm lift}$, the values of 
$T$ and the F-term  $F^T$ are obtained as 
\begin{eqnarray}
k_hT &\simeq& (\ell_0 -\ell_h) S_0, \\
\frac{F^T}{T + T^*} &\simeq& \frac{\ell_0}{\ell_0 - \ell_h}
\frac{m_{3/2}}{\ln (M_P/m_{3/2})}.
\end{eqnarray}
When $\ell_0/(\ell_0 - \ell_h) =O(1)$, 
$F^T/(T + T^*)$ is smaller than $m_{3/2}$ by a factor 
of $O(4 \pi^2)$.
That implies that the modulus mediation due to $F^T$ 
and the anomaly mediation are comparable in this scenario 
\cite{Choi:2005ge,Choi:2005uz}.

We have put anti D3-brane at the tip of warp throat.
Instead of that, a similar uplifting can be realized by 
adding a spontaneous SUSY breaking sector, i.e. 
the F-term uplifting \cite{fuplifting,fuplifting-2}.

\subsection{Soft SUSY breaking terms in the visible sector}

Now, we consider soft SUSY breaking terms in 
the visible sector.
We assume that the compactification scale is close to the 
GUT scale.
Thus, the following initial values are obtained at 
the GUT scale.

First, we study the gaugino masses of the visible gauge sector, 
i.e. $SU(3) \times SU(2) \times U(1)_Y$.
Here, these gauge groups are denoted by $G_a$, ($a=1,2,3$), i.e. 
$G_1=U(1)_Y$, $G_2=SU(2)$ and $G_3=SU(3)$.
Suppose that the SU(3) and SU(2) gauge kinetic functions 
are given as 
\begin{equation}
f_v = T + \ell S,
\label{f-visible}
\end{equation}
 where $\ell$ is a rational number.
The gauge coupling unification in the MSSM, $g^{-2}_{\rm GUT} \simeq
2$, requires ${\rm Re}(T) + \ell {\rm Re}(S_0) \simeq 2$.
The modulus-mediated contributes to the gluino and wino masses 
are obtained as 
\begin{equation}
M_0 = F^T \partial_T \ln ({\rm Re}(f_v))= 
\frac{F^T}{T + T^*}\left( 
\frac{\ell_0 - \ell_h}{\ell_0 - \ell_h + k_h  \ell} \right).
\end{equation}
Since $F^T/(T + T^*)=O(m_{3/2}/(4\pi^2))$, the contributions 
due to the anomaly mediation are comparable.
Thus, just below $M_{GUT}$ the gluino mass $M_3$ and wino mass $M_2$ 
are obtained as 
\begin{equation}
M_a = M_0 + \frac{b_a}{16 \pi^2}g_{GUT}^2m_{3/2},
\end{equation}
with $b_a=1,-3$ for $a=2,3$.
Then, at the energy scale $Q$, these gaugino masses are 
given as 
\begin{equation}
M_a(Q)=M_0\left[ 1 - \frac{1}{8 \pi^2}b_a g_a^2(Q) \ln \left( 
M_{\rm mir}/Q \right) \right] ,
\end{equation}
where the so-called mirage scale $M_{\rm mir}$ is defined as 
\begin{equation}
M_{\rm mir} = \frac{M_{GUT}}{(M_P/m_{3/2})^{\alpha/2}},
\end{equation}
with 
\begin{equation}
\alpha = \frac{m_{3/2}}{M_0 \ln (M_P/m_{3/2})} = 
\frac{\ell_0 - \ell_h + k_h \ell}{\ell_0}.
\end{equation}
When $\alpha =2$, we have $M_{\rm mir} \sim 1$ TeV, that is, 
the gluino and wino masses are unified 
around 1 TeV.
Note that there is no physical threshold around $M_{\rm mir}$.
Here we consider the model with $\alpha =2$.

If we consider the same gauge kinetic function for the $U(1)_Y$ 
group as Eq.~(\ref{f-visible}), the bino mass is also unified 
at $M_{\rm mir}$.
However, the degeneracy between the bino and gluino masses 
is less important to reduce the degree of fine-tuning 
in the Higgs sector, although the degeneracy between 
the wino and gluino masses are important \cite{Abe:2007kf}.
Hence, we consider a generic case for the gauge kinetic 
function of the $U(1)_Y$ group as 
\begin{equation}
f_Y = k_Y T + \ell_Y S,
\end{equation}
in the $U(1)_Y$ charge normalization, which can be embedded 
into the $SU(5)$ GUT.
We assume that  $k_Y {\rm Re}(T) + \ell_Y {\rm Re}(S) \simeq 2$, 
because of the gauge coupling unification, $g^{-2}_{GUT} \simeq 2$.
Then, the modulus-mediated contribution to the bino mass is 
obtained as $k_YM_0$.
The bino mass also has a contribution due to the anomaly 
mediation, and at $M_{GUT}$ the bino mass is obtained as 
\begin{equation}
M_1 = k_Y M_0 + \frac{b_1}{16 \pi^2}g^2_{GUT}m_{3/2},
\end{equation}
where $b_1=33/5$.
Obviously the bino mass $M_1$ 
is not degenerate at $M_{\rm mir}$ unless $k_Y =1$.

Next, we consider soft SUSY breaking scalar masses as well as 
A-terms.
Such SUSY breaking terms are determined by the kinetic term 
of chiral superfield $\Phi^i$,
\begin{equation}
\int d^4 \theta C C^* e^{-K_0/3}Z_i \Phi^{i*}\Phi^i,
\end{equation}
where $Z_i$ is the K\"ahler metric of the matter field $\Phi^i$.
Here, $C$ denotes the chiral compensator superfield, i.e. 
$C=C_0 + F^C \theta^2$, and its F-component is obtained as 
$F^C/C_0=m^*_{3/2}$ in our model. 
Then, the modulus-mediated contributions to A-terms 
and soft scalar masses are obtained as 
\begin{eqnarray}
\tilde A_{ijk} &=& a_{ijk}M_0 = F^T \partial_T \ln (e^{K_0}Z_iZ_jZ_k), \\
\tilde m^2_i &=& c_iM^2_0 = - |F^T|^2 \partial_T \partial_{\bar T} 
\ln(e^{-K_0/3}Z_i),
\end{eqnarray}
where we have assumed that holomorphic Yukawa couplings are 
independent of the modulus $T$.
Here we take the following form,
\begin{equation}
e^{-K_0/3}Z_i=(T + T^*)^{n_i},
\end{equation}
where $n_i$ is a rational number.
Then, $a_{ijk}$ and $c_i$ are obtained as 
\begin{eqnarray}
a_{ijk} &=& (n_i+n_j+n_k)\left( 
\frac{\ell_0 - \ell_h + k_h \ell_h}{\ell_0 - \ell_h} \right), \\
c_i &=& n_i \left( 
\frac{\ell_0 - \ell_h + k_h \ell_h}{\ell_0 - \ell_h} \right)^2.
\end{eqnarray}
A-terms and soft scalar masses have contributions due to 
the anomaly mediation.
Thus, these values at $M_{GUT}$ are given as 
\begin{eqnarray}
A_{ijk} &=& \tilde A_{ijk} - \frac{1}{16\pi^2}(\gamma_i +
\gamma_j + \gamma_k)m_{3/2}, \\
m^2_i &=& \tilde m^2_i-\frac{1}{32\pi^2}\frac{d\gamma_i}{d \ln Q} 
m_{3/2}^2 + \left[ \sum_{jk}\frac{1}{4}|y_{ijk}|^2\tilde A_{ijk} - 
\sum_a g^2_a C^a_2(\Phi^i)r_aM_0 \right] m_{3/2},
\end{eqnarray}
with $r_{2,3}=1$ and $r_1=k_Y$, where $\gamma_i$ denotes the anomalous 
dimension of $\Phi^i$ and $y_{ijk}$ is Yukawa couplings.
In addition, $C^a_2(\Phi^i)$ denotes quadratic Casimir 
of the field $\Phi^i$ under the gauge group $G_a$.
We include RG effects to obtain A-terms and soft scalar masses 
at the energy scale $Q$.
If $k_Y=1$ and the following relations  
\begin{equation}
a_{ijk} = c_i+c_j+c_k=1,
\end{equation}
are satisfied for large Yukawa couplings $y_{ijk}$, 
RG effects and the anomaly mediation contributions cancel 
each other at $M_{\rm mir}$.
Then, we have $A_{ijk}(M_{\rm mir})=\tilde A_{ijk}$ and 
$m^2_i(M_{\rm mir}) = \tilde m^2_i$.
Even for $k_Y \neq 1$, this spectrum is realized approximately 
because RG effects due to $U(1)_Y$ are not important except 
for right-handed slepton masses unless $|k_Y| \geq O(1)$.
We take 
\begin{equation}
c_{H_u}=0, \qquad c_{t_L}=c_{t_R}=\frac{1}{2}.
\end{equation}
Then we can realize the little hierarchy between 
$m_{H_u}=O(M^2_0/(4\pi^2))$ 
and $m_{\tilde t}^2=M_0^2/2$.
We consider $M_0 =O(1)$ TeV and a moderate value of $\tan \beta$, 
e.g. $\tan \beta=10$.
Then, we neglect all of Yukawa couplings except the top Yukawa 
coupling.
For the down-sector Higgs field, we take $c_{H_d}=1/2$.

With the above assignment of $a_{ijk}$ and $c_i$, 
we have a smaller higgsino mass $\mu=100 - 200$ GeV.
Thus, a small value of $|k_Y|$, $|k_Y| < 1$, is interesting 
because in such a case the LSP would be a mixture between 
the higgsino and bino.
If $k_Y$ and the bino mass are quite small, 
right-handed slepton masses for $c_{\ell}=1/2$ 
may become tachyonic at the weak scale.
Thus, we take $c_{\ell}=1$ for both left-handed and right-handed 
slepton masses.
At any rate, slepton masses are irrelevant to the fine-tuning 
problem of the Higgs sector.

Alternatively, in order to increase slepton masses 
we could consider the scenario with 
an extra (anomalous) $U(1)$ gauge group.
We assume that such $U(1)$ sector is separated away from the 
SUSY breaking anti D3 brane, and $U(1)$ is broken 
at a certain scale, e.g. $M_{GUT}$.
Such breaking induces another source of soft scalar masses, which 
are proportional to $U(1)$ charge $q_i$ of the fields $\Phi^i$, 
\begin{equation}
m^2_{i(D)}= q_i D.
\end{equation}
The size of $D$ is model-dependent.\footnote{See 
Ref.~\cite{Choi:2006bh,Brummer:2006dg} for the D-term 
contributions in the KKLT scenario 
and Ref.~\cite{Kawamura:1996wn,Higaki:2003ig} for the D-term 
in dilaton-moduli mediation of heterotic string and type I string 
theory.}
This type of contribution could also increase 
slepton masses.

The size of sleton masses are important for analysis 
on the thermal relic density of the LSP as shown in the next section.
Hence, in the following section we consider two cases for slepton 
masses, 1) the case that slepton masses are determined from
$c_{\ell}=1$ and 2) the case that slepton masses vary.
The latter case can be realized by the D-term contributions.

\section{Neutralino Dark Matter}

In this section, we consider neutralino dark matter phenomenology.
Recent WMAP and other observations imply that the cold dark matter abundance is \cite{wmap3}
\begin{eqnarray}
0.085 < \Omega_{\rm DM} h^2 <0.119~ (95\%~ \rm CL),
\label{wmap}
\end{eqnarray}
where $h \simeq 0.7$ is the scaled Hubble constant.
We assume that the neutralino LSPs were in thermal equilibrium when the temperature
of the Universe is larger than the LSP mass $m_\chi$. As the temperature drops below $m_\chi$,
the number density of the LSP is exponentially suppressed. At some point neutralino LSP
annihilation rate becomes smaller than the Hubble expansion rate. 
Then the neutralino LSPs fall out of equilibrium and the LSP number density in a comoving volume 
remains constant \cite{jungman}.
We also assume that the neutralino LSP constitutes 
all the cold dark matter in the Universe at the current epoch.

\begin{figure}[ht!]
\vskip 0.6cm
\begin{center}
\epsfig{figure=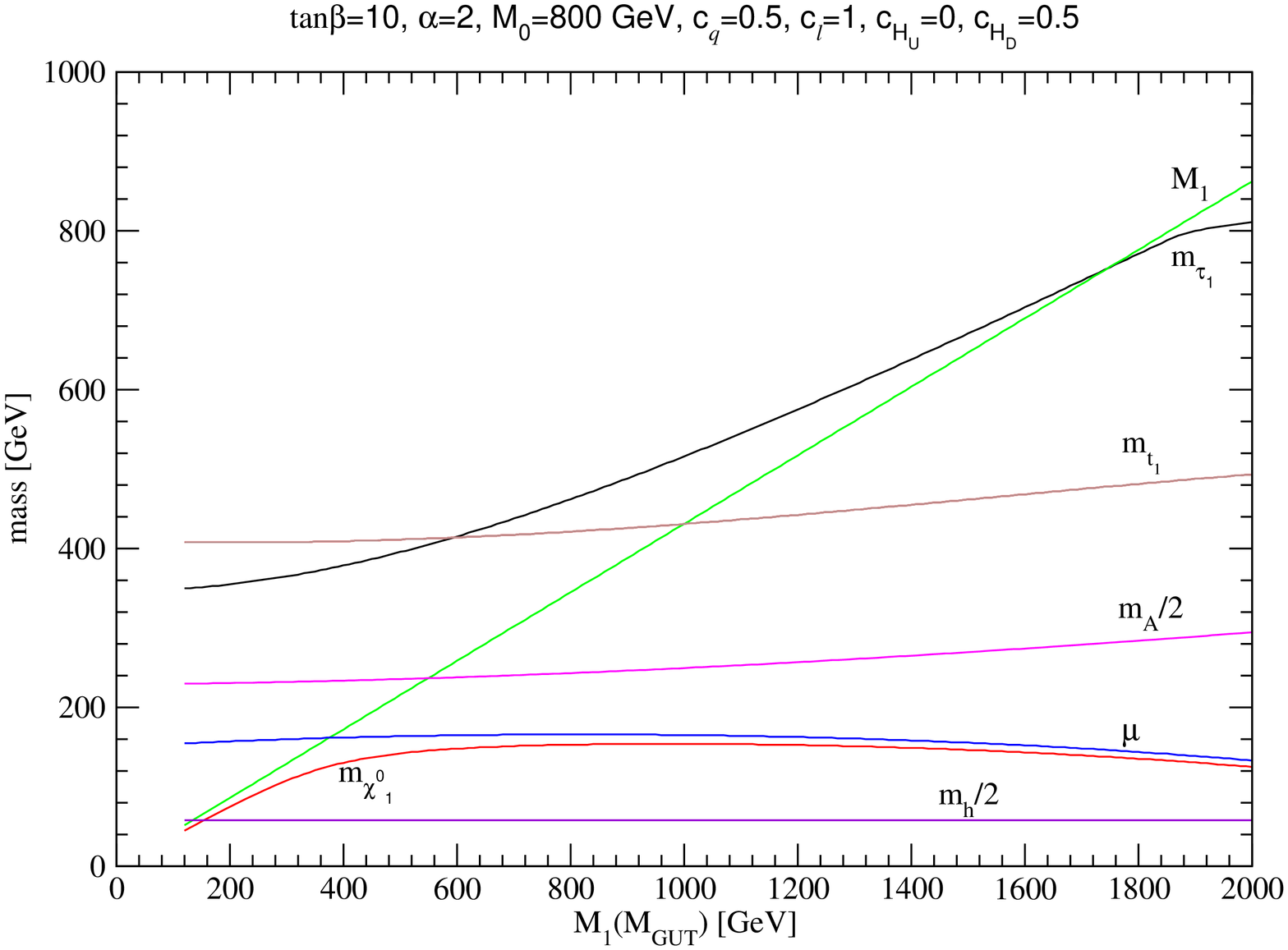,width=7cm,height=7cm,angle=270}\hskip 0.5cm
\epsfig{figure=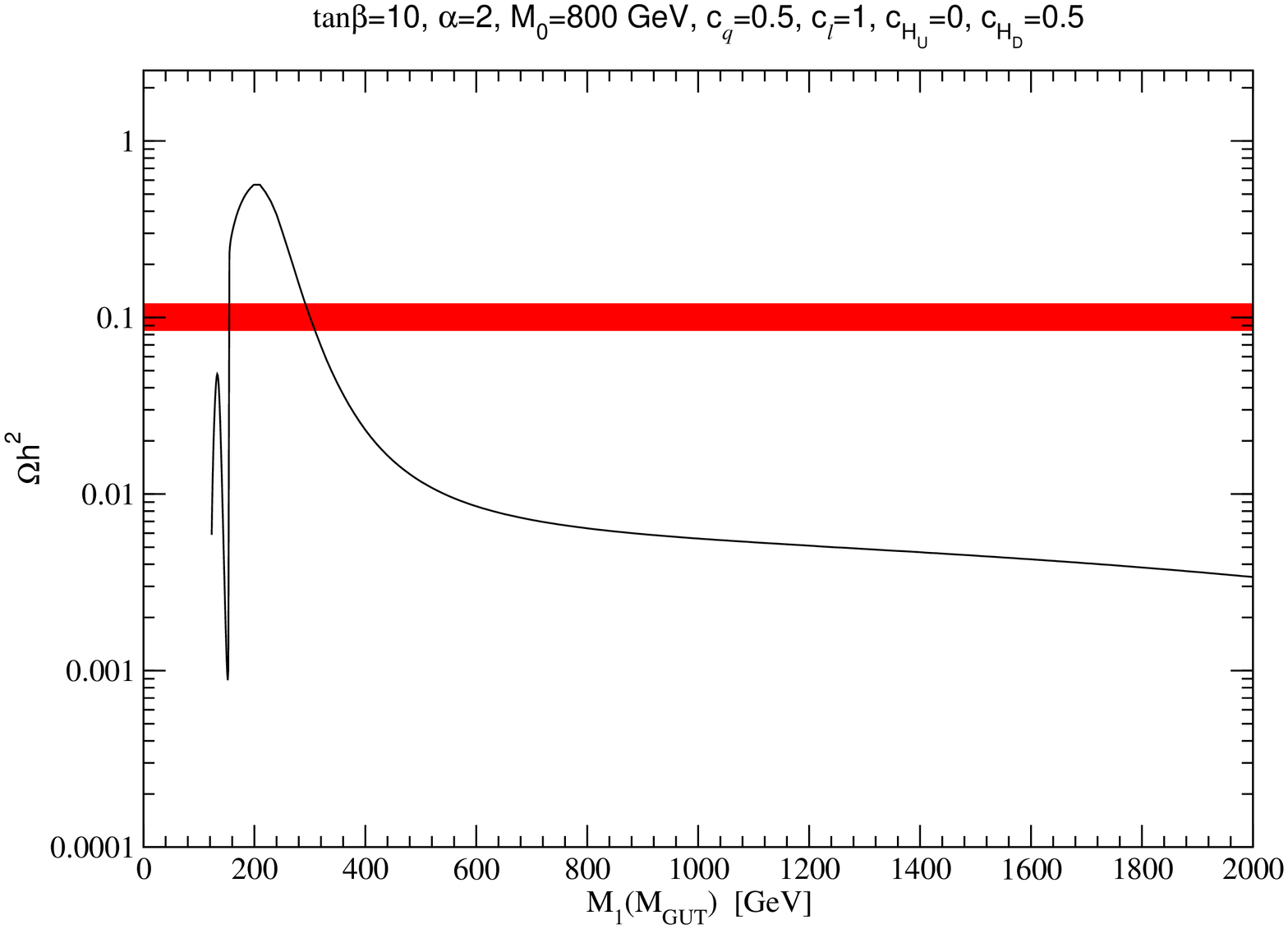,width=7cm,height=7cm,angle=270}
\end{center}
\vskip -0.4cm
\caption{\it (a) sparticle masses at the weak scale and (b) thermal relic density $\Omega_\chi h^2$
of neutralino LSP, as a function of $M_1 (M_{\rm GUT})$.}
\label{fig:oh2}
\end{figure}

In the TeV scale mirage mediation, in which all three gaugino masses are unified at TeV scale,
it turns out that the neutralino LSP is higgsino-like \cite{Choi:2006im}.
This is because the gluino mass $M_3$ is smaller than bino mass $M_1$ (and wino mass $M_2$)
at higher energy scales. (Notice that $M_3: M_2: M_1 \simeq (1-0.3\alpha) g_3^2:
(1+0.1\alpha) g_2^2: (1+ 0.66\alpha) g_1^2$, and $\alpha \sim 2$ for 
the TeV scale mirage mediation.)
Such a small $M_3$ gives a small stop mass squared and in turn 
leads to a small $|m_{H_u}^2|$ and thus $|\mu|$ at the weak scale, 
compared to bino mass $M_1$ and wino mass $M_2$.

For higgsino-like LSP, the lighter chargino $\chi_1^\pm$ and the two light neutralinos
$\chi_1^0$, $\chi_2^0$ are nearly degenerate. In this case 
dominant annihilation processes for the neutralinos and chargino
are neutralino pair annihilation into gauge bosons, and 
the neutralino-neutralino and neutralino-chargino coannihilations 
into fermion pair \cite{coanil}.
These annihilation processes are so effective that the thermal relic density of
the neutralino LSP is too small unless neutralino LSP is rather heavy 
($m_{\chi_1^0} \sim |\mu| \sim 1$ TeV).
Therefore, the higgsino LSP with $|\mu| \sim O(M_Z)$, which avoids fine-tuning
in the Higgs sector, cannot provide the correct amount of thermal
relic density in Eq.(\ref{wmap}).

On the other hand, the TeV scale partial mirage unification scenario can provide 
not only a solution of the little hierarchy problem  but also
a right amount of thermal relic density. 
As a specific numerical example, we choose a parameter set; 
\begin{eqnarray}
\alpha=2,\  M_0=800~ {\rm GeV},\
c_{H_u}=0,\ c_{H_d}=1/2,\  c_q=1/2,\ c_l=1,\ \rm tan\beta=10,
\label{param1}
\end{eqnarray}
while varying the bino mass at the GUT scale within some range, 100 GeV $< M_1 (M_{\rm GUT}) <$ 2 TeV.
Figures $\ref{fig:oh2}$ show (a) sparticle masses at the weak scale and
(b) thermal relic density $\Omega_\chi h^2$ of the neutralino LSP, as a function of $M_1 (M_{\rm GUT})$.
One can notice that $\mu$ values at weak scale remain small $i.e,$
$130 \lesssim \mu \lesssim 160$ GeV so that there is no little hierarchy problem in this case.
Our model leads to the CP even Higgs mass, $m_h \sim 116$ GeV.

For a large $M_1 (M_{\rm GUT})$ value, $\mu$ is much smaller than
$M_1$ at the weak scale implying higgsino-like
LSP. It leads to a very small relic density $\Omega_\chi h^2 \sim O(10^{-3})$.
The bino mass at the weak scale decreases as $M_1 (M_{\rm GUT})$ decreases,
and becomes similar to $\mu$ value at the weak scale when $M_1 (M_{\rm GUT}) \sim 350$ GeV.
In the bino-higgsino mixed region of LSP, 
the relic density $\Omega_\chi h^2$ increases rapidly
as $M_1 (M_{\rm GUT})$ decreases, due to the enhanced bino-component of neutralino LSP.
When $M_1 (M_{\rm GUT}) \sim 300$ GeV, $\Omega_\chi h^2 \simeq 0.1$, thus providing
a right amount of relic density which is consistent with the WMAP bound
on the dark matter density.

As $M_1 (M_{\rm GUT})$ further decreases, the neutralino LSP becomes bino-like and the relic density  
$\Omega_\chi h^2$ gets too large and increases until $M_1 (M_{\rm GUT}) \sim 200$ GeV. 
Below this point, an interesting annihilation channel for neutralino LSP is open.
For the region around $M_1 (M_{\rm GUT}) \simeq 160$ GeV, the mass of the neutralino LSP is equal to
the half of the light CP even higgs mass, $i.e, m_{\chi_1^0} \sim m_h /2$.
In this case the neutralino pair annihilation through s-channel higgs exchange becomes 
very efficient so that the relic density $\Omega_\chi h^2$ is reduced to a very small values $O(10^{-3})$, 
passing acceptable ones $O(10^{-1})$. When $m_{\chi_1^0} \sim m_Z /2$, $Z$ resonance effect is dominant
for reducing the relic density.

\begin{figure}[ht!]
\vskip 0.6cm
\begin{center}
\epsfig{figure=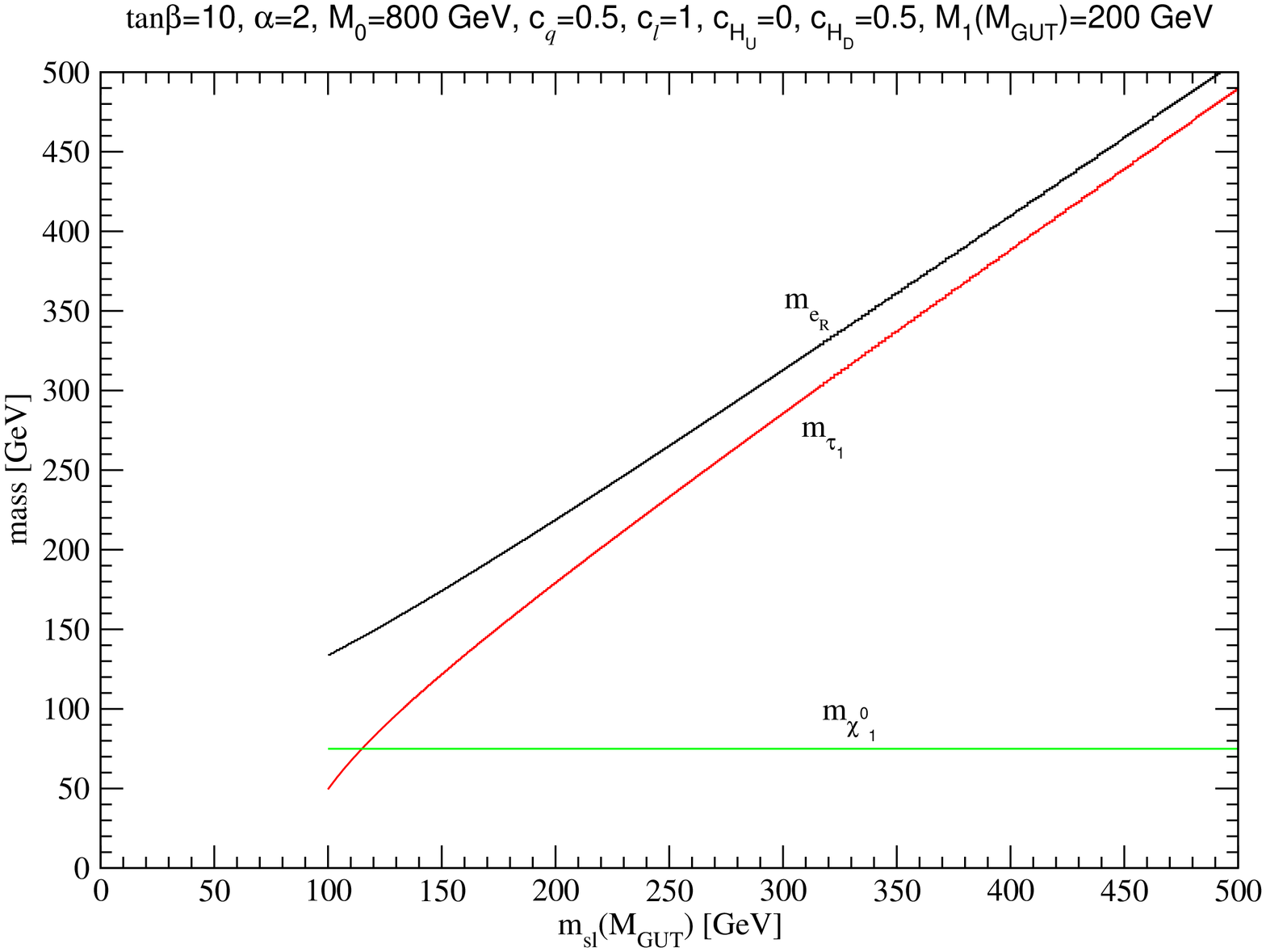,width=7cm,height=7cm,angle=270}\hskip 0.5cm
\epsfig{figure=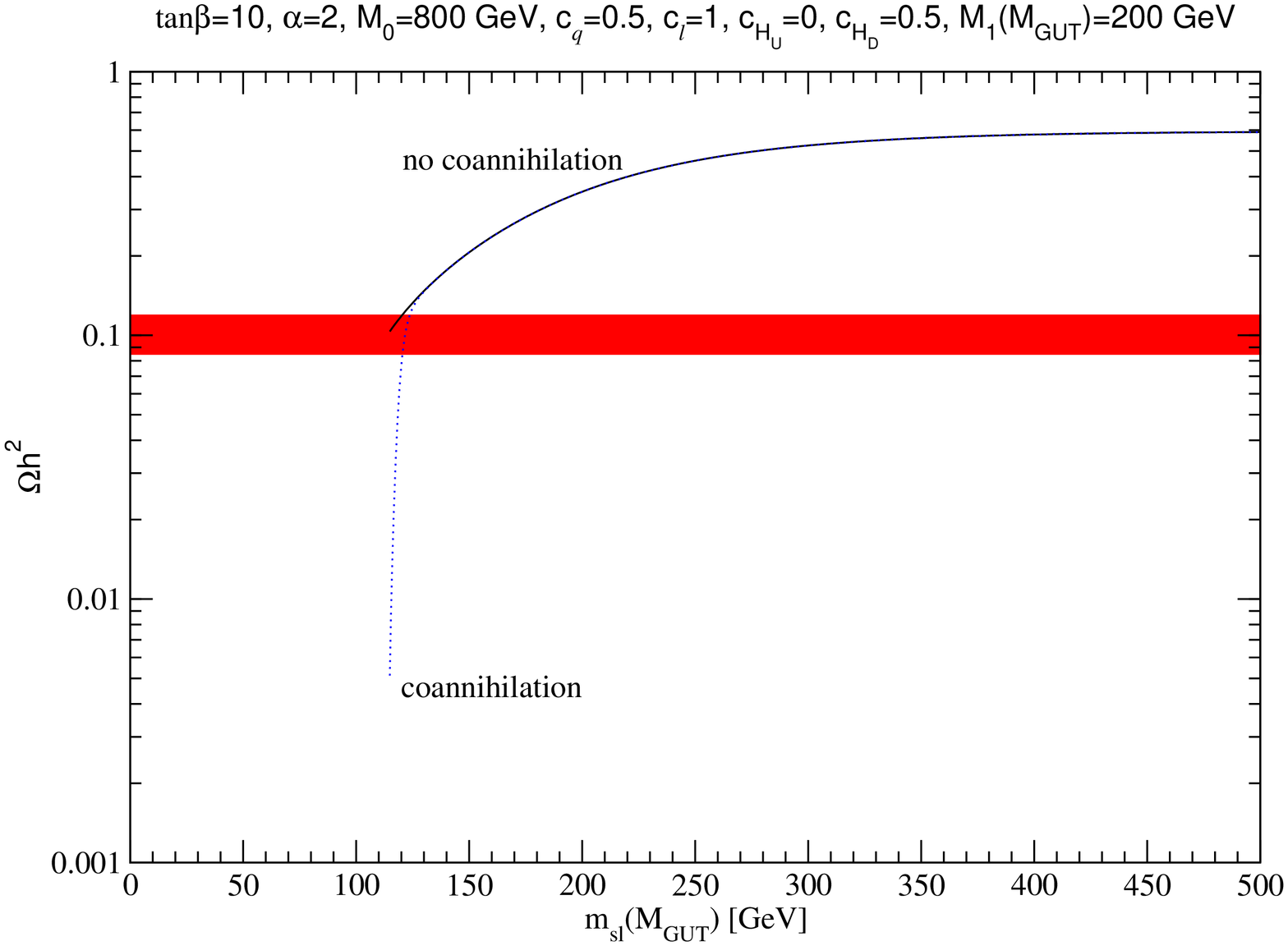,width=7cm,height=7cm,angle=270}
\end{center}
\vskip -0.4cm
\caption{\it (a) slepton masses at the weak scale and (b) thermal relic density $\Omega_\chi h^2$
of neutralino LSP, as a function of $m_{slepton}$(GUT).}
\label{fig:slepton}
\end{figure}

As we discussed in the last section, for the scenario with an extra (anomalous) $U(1)$ gauge group,
additional $D$-term contributions to soft terms would make slepton mass a free parameter in practice.
In order to see possible effects of the $D$-term contributions, we fix $M_1 (M_{\rm GUT})=200$ GeV 
with the parameter set $(\ref{param1})$ while
varying $m_{\rm sl} (M_{\rm GUT})$, the slepton mass at the GUT scale. 
Notice that in this case, the neutralino LSP is bino-like and $m_{\chi_1^0} \simeq 75$ GeV.
Figures 2 show (a) slepton masses at weak scale and (b) the thermal relic density 
$\Omega_\chi h^2$, as a function of $m_{\rm sl} (M_{\rm GUT})$. 
When the slepton mass is large ($m_{e_R} \simeq 500$ GeV), 
the relic density is quite large ($\Omega_\chi h^2 \simeq 0.6$), as expected for the bino-like LSP
with rather heavy sparticle mass spectrum.
As $m_{\rm sl}(M_{\rm GUT})$ decreases, however, slepton masses at weak scale decrease. 
Accordingly, the relic density $\Omega_\chi h^2$ decreases and gets close to 
the WMAP bound (\ref{wmap}) when $m_{e_R} \sim 150$ GeV. 
It is known that in this case, the LSP relic density is mainly determined 
from neutralino pair annihilation into lepton pair through $t$-channel exchange of SU(2) singlet 
sleptons \cite{dn}. 

For $m_{\rm sl} (M_{\rm GUT}) \lesssim 115$ GeV, 
the neutralino LSP and the lighter stau are almost degenerate. 
Then LSP-stau coannihilation \cite{ellis}
becomes very effective to reduce thermal relic density of the neutralino LSP.
>From the Fig. 2(b), one can notice that
the thermal relic density $\Omega_\chi h^2$ reaches 
the WMAP range and then drops quickly below 0.01
in the small $m_{\rm sl} (M_{\rm GUT})$ region.

\begin{figure}[ht!]
\vskip 0.6cm
\begin{center}
\epsfig{figure=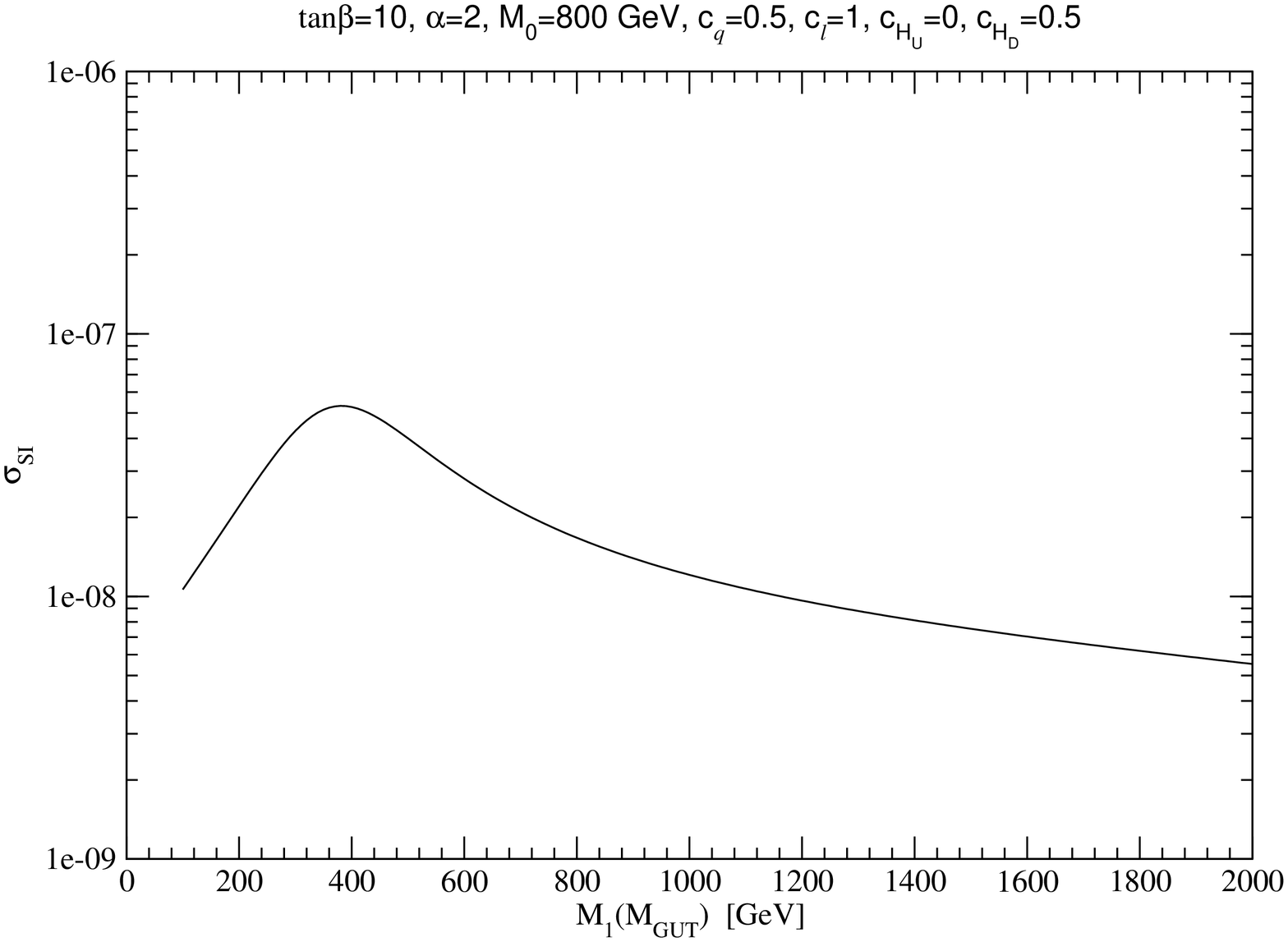,width=7cm,height=7cm,angle=270}\hskip 0.5cm
\epsfig{figure=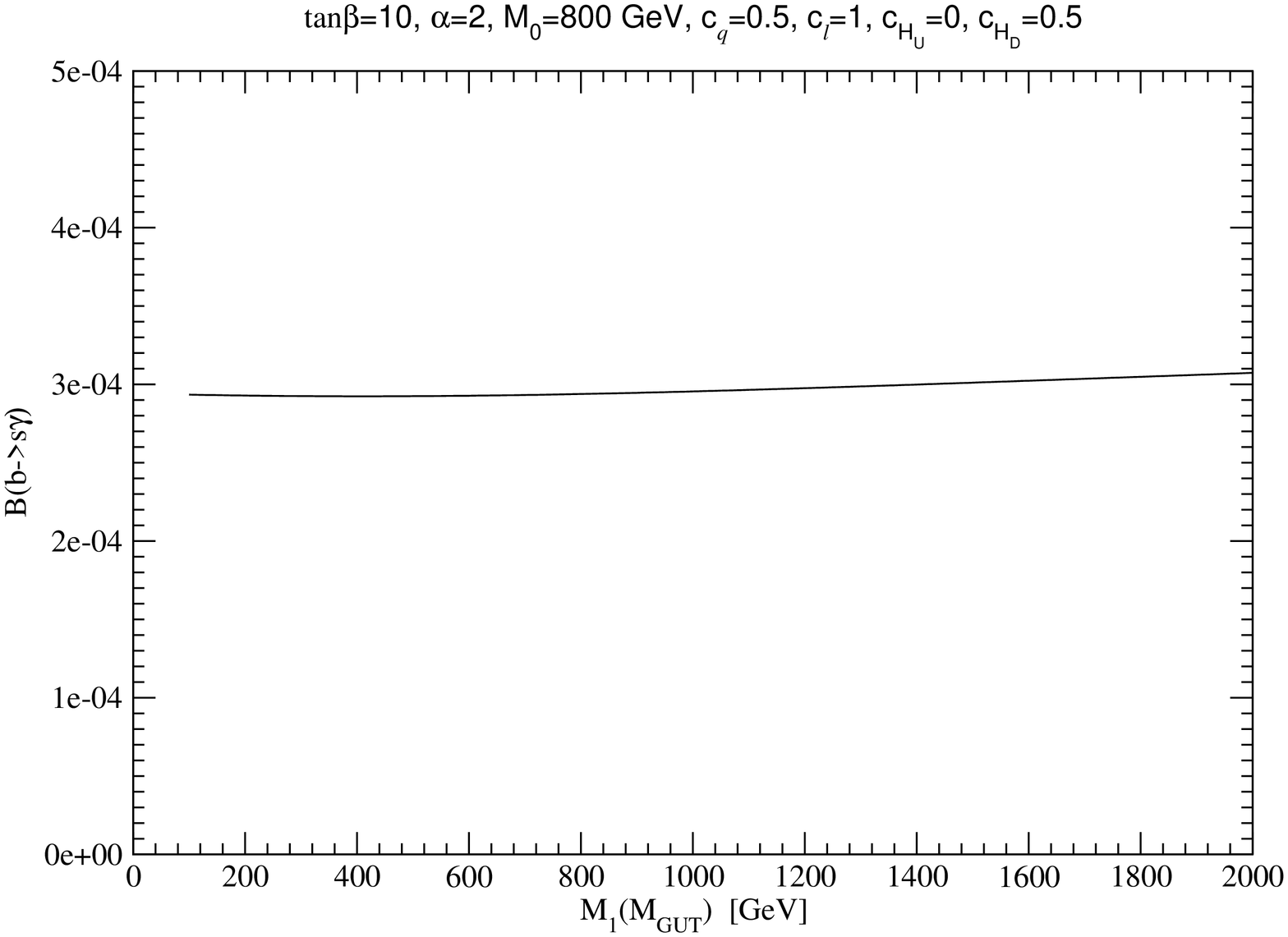,width=7cm,height=7cm,angle=270}
\end{center}
\vskip -0.4cm
\caption{\it (a) spin-independent cross section of neutralino and proton 
and (b) $b\rightarrow s\gamma$ branching ratio, as a function of $M_1(M_{\rm GUT})$.}
\label{fig:scattering}
\end{figure}

Our model has also an interesting aspect for the direct detection search of 
neutralino dark matter. For the spin-independent cross section of neutralino-proton 
scattering, the contributions from $t$-channel CP even Higgs exchanges are usually dominant \cite{jungman}. 
The cross section $\sigma_{SI}$ would enhance if neutralino LSP is a mixed state of gaugino and higgsino,
due to the nature of neutralino-neutralino-Higgs couplings. 
Figure \ref{fig:scattering}(a) shows the spin-independent 
scattering cross section as a function of $M_1(M_{\rm GUT})$ with the parameters (\ref{param1}). 
The $\sigma_{SI}$ is quite small ($\sim 5\times 10^{-9}\ \rm pb)$ for large $M_1(M_{\rm GUT})$ region $i.e$,
higgsino-like LSP case. There is, however, about one order of
magnitude increase of $\sigma_{SI}$ in
the bino-higgsino mixed region. For $M_1(M_{\rm GUT}) \simeq 300$ GeV, which provides 
a right amount of thermal relic density $\Omega_\chi h^2 \simeq 0.1$,
the spin-independent scattering cross section is $\sigma_{SI} \simeq 4\times 10^{-8}\ \rm pb$ with
$m_{\tilde\chi_1^0} \simeq 105$ GeV. This cross section value is quite close to the current limit from 
XENON experiment \cite{xenon}, $i.e.,\ 8.8\times 10^{-8}$ pb for a WIMP mass of 100 GeV.
Therefore, our model would be explored in the near future experiments on the direct searches.

Before closing this section, we comment on experimental constraints.
Our benchmark point (\ref{param1}) satisfy the experimental bounds on particle masses such as
$m_{\chi^+} > 104$ GeV for chargino and $m_h > 114$ GeV for light Higgs boson.
Figure \ref{fig:scattering}(b) shows $b\rightarrow s\gamma$ branching ratio $BR(b\rightarrow s\gamma)$ 
for the parameter choice (\ref{param1}), as a function of $M_1(M_{\rm GUT})$. 
The NLO calculation for $BR(b\rightarrow s\gamma)$ gives about $3\times 10^{-4}$ for our 
parameter choice, insensitive to $M_1(M_{\rm GUT})$.
These predictions are rather smaller than the current world average of experimental values \cite{bsg}, 
$BR(b\rightarrow s\gamma)^{exp}=(3.55\pm 0.26)\times 10^{-4}$, 
due to large contribution from chargino-stop loop which adds destructively to 
Standard Model contribution for our choice on the sign of $\mu\ (> 0)$.
Considering theoretical and experimental uncertainties, it turns out that the calculated branching ratio is 
consistent with the measured one within $2\sigma$ range.

\section{Conclusions}

We have studied the TeV scale partial mirage unification scenario, 
where the gluino and wino masses are degenerate, but the bino mass 
is not degenerate.
We have shown an example leading to such a spectrum.
This spectrum has phenomenologically interesting aspects.
First, there is no fine tuning problem because of 
the degeneracy of the gluino and wino masses, that is, 
our model leads to $130 \lesssim \mu \lesssim 160$ GeV.
The LSP is the mixture of the bino and higgsino. 
In the TeV scale partial mirage unification, 
a right amount of thermal relic density of neutralino LSP
can be obtained through various channels for neutralino annihilations.
A mixed bino-higgsino LSP, which is available through adjusting the bino mass at the GUT scale, 
may lead to an appropriate neutralino annihilation rate into gauge bosons 
and so the right amount of the relic density. 
The neutralino pair annihilation via s-channel higgs exchange
play an important role for obtaining the suitable relic density, when $m_\chi \sim m_h/2$ 
in bino-like LSP region. Furthermore, if the slepton mass can vary independently, 
LSP annihilations through $t$-channel SU(2) singlet slepton exchange or LSP-stau coannihilation
can make the thermal relic density satisfy the WMAP bound on dark matter density.
The TeV scale partial mirage unification scenario also provides a sizable 
spin-independent scattering cross section between neutralino dark matter and nucleon, 
which can be explored in near future experiments, when the neutralino dark matter is
a mixture of bino and higgsino.

\subsection*{Acknowledgement}

H.~A.\/ and T.~K.\/ are supported in part by the Grand-in-Aid for 
Scientific Research \#182496, \#17540251, respectively. 
T.~K.\/ is also supported in part by the Grant-in-Aid for the 21st Century 
COE ``The Center for Diversity and Universality in Physics'' from the 
Ministry of Education, Culture, Sports, Science and Technology of Japan. 
This work was supported by the KRF Grant KRF-2005-210-C000006 funded by the Korean Government and
the Grant No. R01-2005-000-10404-0 from the Basic Research Program of the Korea
Science \& Engineering Foundation (Y.G.K. and Y.S.).

\end{document}